\documentclass[journal]{IEEEtran}
\usepackage{amsmath, amssymb, bbm}
\usepackage{bm}
\usepackage{amsthm}
\usepackage{graphicx,epsfig}
\usepackage{subfigure}
\usepackage{verbatim}
\usepackage{soul}
\usepackage[colorlinks]{hyperref}
\usepackage{color}
\usepackage{booktabs}
\usepackage{makecell}
\usepackage{graphicx}
\usepackage{enumitem}
\usepackage{color}

\usepackage{algorithm}
\usepackage{algpseudocode}

\newtheorem*{proof*}{Proof}

\usepackage{hhline}
\usepackage{multirow}
\usepackage{caption}

\usepackage{cite}

\usepackage[utf8]{inputenc}
\usepackage{amssymb}
\usepackage{nomencl}
\usepackage{etoolbox}

\begin{document}
\title{\Large{A Gaussian Process-based Price-Amount Curve Construction for Demand Response Provided by Internet Data Centers}}
\author{Yang Liu,\textit{ Member}, \textit{IEEE} and Hung D. Nguyen, \textit{Member}, \textit{IEEE}

\thanks{ Yang Liu and Hung D. Nguyen, e-mail: \{liu-yang, hunghtd\}@ntu.edu.sg.

\textsuperscript{\,} Corresponding author: Hung D. Nguyen}}

\markboth{IEEE Transactions on Smart Grids,~Vol.~, No.~, }{}%

\maketitle

\begin{abstract}
For a Demand Response (DR) program with internet data centers (IDC), the Price-Amount curve that estimates how the potential DR amount depends on the DR price determined by power systems is crucial. Constructing this curve is challenging mainly due to the uncertainty in IDCs' operation. A novel Gaussian Process Regression-based estimation method is thus proposed. The variance of resulting curve reflecting the IDC operational uncertainty is also calculated. 
\end{abstract}


\section{Introduction} 
With the blooming of IT industry, Internet Data Centers (IDCs) become an integral part of modern power systems. Compared to the conventional loads, the demand of IDCs is quite flexible not only because IDCs are usually equiped with Energy Storage Systems (ESSs) and renewable distributed generators (DGs), but also the workloads in IDCs can be migrated temporally and spatially. The flexible demand makes it possible for IDCs to participate in electricity markets, such as demand response (DR) programs. Such DR programs may require IDCs to provide a bidding curve about the DR price and the provided DR amount. Due to the instantaneity of IDC workloads and the uncertainties in IDC operation, the bidding curve built empirically based on historical data might not be reliable. While IDCs in electricity markets has been considered in the literature \cite{idc1,IDC3}, incorporating IDC operation in constructing this curve has not been sufficiently developed. 

To handle instant workloads and operational uncertainties, a Gaussian Process Regression (GPR)-based method is proposed to estimate the DR Price-Amount function. By sampling the IDC operation states in the estimation horizon, the relationship between the DR price and the provided DR amount under uncertainties appears and is captured by GPR. With this relationship, power systems can not only adjust the DR amount as needed, but also foreseen the uncertain level of the actual available DR amount, which is very useful for demand-side management. The numerical results show the effectiveness and accuracy of the proposed method. 

\section{Data Center Operation Model}

IDC operation consists of two parts: the power level operation of energy flow in IDCs as a microgird with load, ESS and DGs; the server level operation of workload adjustment, which determines the load in the power level. Commonly, the objective of IDC operation is to minimize the operational cost, including the workload termination cost and the electricity bill, minus the revenue obtained from DR programs. Considering the uncertainties in the two levels, robust optimization is widely used in IDC operation \cite{data_center_robust}. 

Equation (\ref{objective}) describes the optimization objective of IDC operation
\begin{equation}
        \min_{\bm{\mathrm{V^{1st}}}} \left\{
    \begin{aligned}
    &\sum_{wl}v^t_{wl}\,p^{WL}_{wl} \\ 
    + &\max_{\bm{\mathrm{V^{2nd}}}} \left\{
        \begin{aligned}
             \sum_{idc,t}P_{idc,t}\,p_{idc,t}-\sum_{idc,t}P_{idc,t}^{DR}\,p_{idc,t}^{DR}\\:\bm{\mathrm{Z}} \in [\bm{\mathrm{\underline{Z}}},\bm{\mathrm{\overline{Z}}}]  
        \end{aligned}
        \right\}
    \end{aligned}
    \right\} \label{objective}
\end{equation}
where $idc$, $t$ and $wl$ is the index of IDCs, time slots, and workloads, respectively;
$p^{WL}$, $p$ and $p^{DR}$ is the price of flexible workloads, electricity (uncertain) and DR, respectively; $v^{t}$ is the binary indicator of workload termination (1 if terminated); $P$ is the power from grids; $P^{DR}$ is the provided DR amount; $\bm{\mathrm{V^{1st}}}$ is the first stage variable set describing server level operation; $\bm{\mathrm{V^{2nd}}}$ is the second stage variable set describing power level operation, which are determined at each time slot after the uncertainties are known; $\bm{\mathrm{Z}}$ is the uncertain parameter set; $\bm{\mathrm{\overline{Z}}}$ and $\bm{\mathrm{\underline{Z}}}$ is the upper and lower boundaries of $\bm{\mathrm{Z}}$ varying range, respectively. Noting that the Real Time Pricing (RTP) DR program is assumed allowing the provided DR amount is determined in very short horizon (such as the next 15 mins) \cite{RTP}, so $P^{DR}$ is defined as a second stage variable . 

The detailed model of IDC operation in the power and server level is described in the following subsections as the constraints of the robust optimization model.

\subsection{Internet Data Center Power Model}
The energy consumption of a IDC mainly includes two parts: the consumption of IT servers and the consumption of other equipment such as cooling systems. Generally, the consumed energy of other equipment is proportional to the IT consumption \cite{PUE}. So the total power demand $P^D$ can be calculated by the IT power demand $P^{IT}$ as follows: 
\begin{equation}
    P^D_{idc,t}=P^{IT}_{idc,t}*PUE_{idc,t} \qquad\qquad\qquad\qquad \forall idc,t \label{eq1}
\end{equation}
where $PUE$ is Power Usage Effectiveness (PUE), the ratio between the IT power demand and the total demand.

Beside extracting power from grids, IDCs can satisfy the demand with the equiped ESS and DGs as a microgrid. The power model of IDCs is formulated in the following equations.
\begin{align}
& P_{idc,t}=P^D_{idc,t}+P^{ESS}_{idc,t}-P^G_{idc,t} &\qquad\qquad \forall idc,t \label{eq2} \\
& 0 \leq P_{idc,t} \leq P^{Nomi}_{idc,t}-P^{DR}_{idc,t} &\quad \forall idc,t \label{eq3}\\
& E^{ESS}_{idc,t}=E^{ESS,Initial}_{idc,t}+\sum_{t'=1}^{t}P^{ESS}_{idc,t'} &\quad \forall idc,t \label{eq4} 
\end{align}
\begin{align}
0 \leq E^{ESS}_{idc,t} \leq E^{ESS,Max}_{idc} &\quad \forall idc,t \label{eq5}
\end{align}

Equation (\ref{eq1}) shows the power balance of IDCs: the power from grids equals to the demand of the IDC plus the ESS charging/discharging power $P^{ESS}$ (positive means charging) minus the DG output $P^G$ (uncertain). For the IDCs participating in DR programs, the power from grids should be less than the nominal power demand $P^{Nomi}$ minus the provided DR amount $P^{DR}$ as shown in (\ref{eq3}). The ESS energy $E^{ESS}$ is modeled in (\ref{eq4}) and (\ref{eq5}), where $E^{ESS,Initial}$ and $E^{ESS,Max}$ is the initial and maximum energy of the ESS. Noting that $E^{ESS}$ and $P^{ESS}$ belong to the first stage variables since they are determined before the operation horizon.

\subsection{Internet Data Center Workload Adjustment Model}
Workloads in IDCs can be roughly categorized to two types: the interactive workload flow and the flexible workloads. The former is like the ordinary demand in power systems: it has to be satisfied in real time and we can only forecast the amount of it; the latter is more like the demand of scheduled tasks: it can be migrated temporally and spatially or even terminated when needed, and the demand of each task is known. Thus, IDCs should dispatch servers to satisfy the interactive workload flow in real time, and determine the migration and termination of flexible workloads in the operation horizon. The power demand on workload computation is modeled in Eq. (\ref{eq6}), where $P^{in}$ is the power demand of interactive workload flow (uncertain); $P^{fl}$ is the demand of a flexible workload. Due to space limits, please refer to our previous work in \cite{8527554} for the rest of workload adjustment model.

\begin{equation}
P_{idc,t}^{IT} = P_{idc,t}^{in} + \sum_{wl}P^{fl}_{wl,idc,t}*(1-v^t_{wl})  \qquad \forall idc,t \label{eq6}
\end{equation}

\section{Gaussian Process Regression based DR Price-Amount Function Estimation}
In this section, the process of using GPR to obtain the function between the DR price and the DR amount provided by IDCs is explained. The key idea of GPR can be summarised as follow: for an unknown function $y=f(x)$ and a sample set $S$ consisting of $m$ pairs of the independent variable $R=[x^{(0)},x^{(1)},...,x^{(m)}]$ and the related dependent variable $D=[\hat{y}^{(0)},\hat{y}^{(1)},...,\hat{y}^{(m)}]$, the estimated output of the function with a given input $x$ will be:
\begin{equation}
    \hat{y}=\hat{f}(x)=k(x,D)^T(k(D,D)+\sigma^2I)^{-1}R
\end{equation}
and the variation of the estimation results at point $x$ due to the uncertainties in the function will be:
\begin{equation}
    \Sigma(x)=k(x,x)+\sigma^2I-k(x,D)(k(D,D)+\sigma^2I)^{-1}k(D,x)
\end{equation}
where $k$ refers to kernel functions. 
The advantages of GPR over other similar methods include: 1) GPR can provide an analytic expression of the estimated function, which is quite important for the further applications based on the function; 2) As a non-parametric method, GPR doesn't have strong assumptions of the function; 3) GPR is a Bayesian method which can provide the estimation confidence level affected by the uncertainties; 4) By selecting a proper kernel, GPR can achieve accurate estimation with limited number of samples.

\begin{figure}[ht]
\centering 
\includegraphics[width=0.45\textwidth]{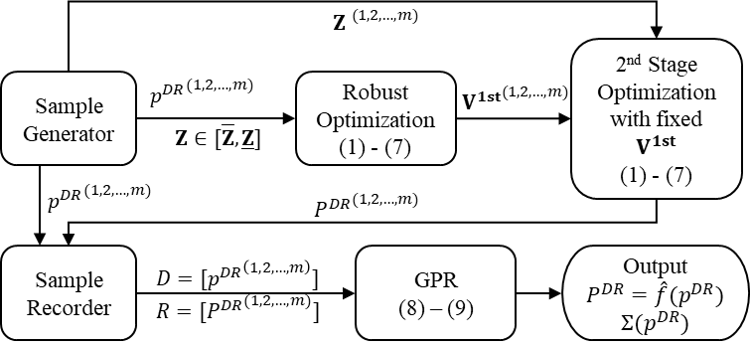} 
\caption{Procedures of Proposed GPR-based Estimation Method} 
\label{GPR} 
\end{figure}

The procedures using GPR to estimate the function between the DR amount provided $P^{DR}$ and the DR price $p^{DR}$ is illustrated in Fig. \ref{GPR}. The possible value of $p^{DR}$ is generated with the GP-UCB based sampling method which can reduce  the  sample  number without compromising accuracy \cite{GP-UCB}, and the generated value is inputed into the IDC robust optimization model with the varying range of the uncertain parameters $\bm{\mathrm{Z}}$ to calculate the first stage variables $\bm{\mathrm{V^{1st}}}$ determining workload adjustment. After $\bm{\mathrm{V^{1st}}}$ is obtained, the value of uncertain parameters is randomly generated according to the probability distribution, and then the second stage optimization is solved to find the optimal DR amount $P^{DR}$. By repeating this process, a sample set consisting of independent variable $D=[{p^{DR}}^{(1)},{p^{DR}}^{(2)},...,{p^{DR}}^{(m)}]$ and dependent variable $R=[{P^{DR}}^{(1)},{P^{DR}}^{(2)},...,{P^{DR}}^{(m)}]$ is generated. Taking the sample set into GPR, the estimated function $P^{DR}=\hat{f}(p^{DR})$ and the estimation variance $\Sigma(p^{DR})$ can be obtained.

\section{Numerical Results}
\begin{figure}[ht]
\centering 
\includegraphics[width=0.45\textwidth]{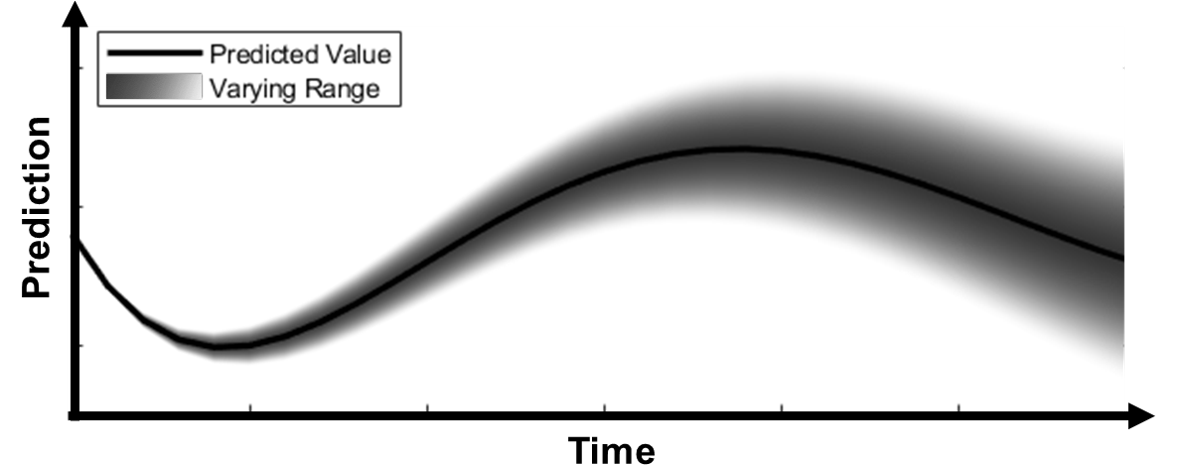} \caption{Prediction Error Increasing with Prediction Horizon} 
\label{prediction} 
\end{figure}

In this section, the proposed method are used to estimate the DR Price-Amount function for two IDCs one hour ahead (four time slots, 15 mins per time slot). One thousand IDC operation samples are generated with the GUROBI solver in a PC with 3.2GHz 8-Core CPU and 16GB RAM. The IDC interactive workload flow, renewable DG output, and electricity price are uncertain parameters with different probability distribution, and the uncertainties brought by prediction errors are assumed increasing with prediction horizon as shown in Fig. \ref{prediction}. 

\begin{figure}[ht]
\centering 
\includegraphics[width=0.45\textwidth]{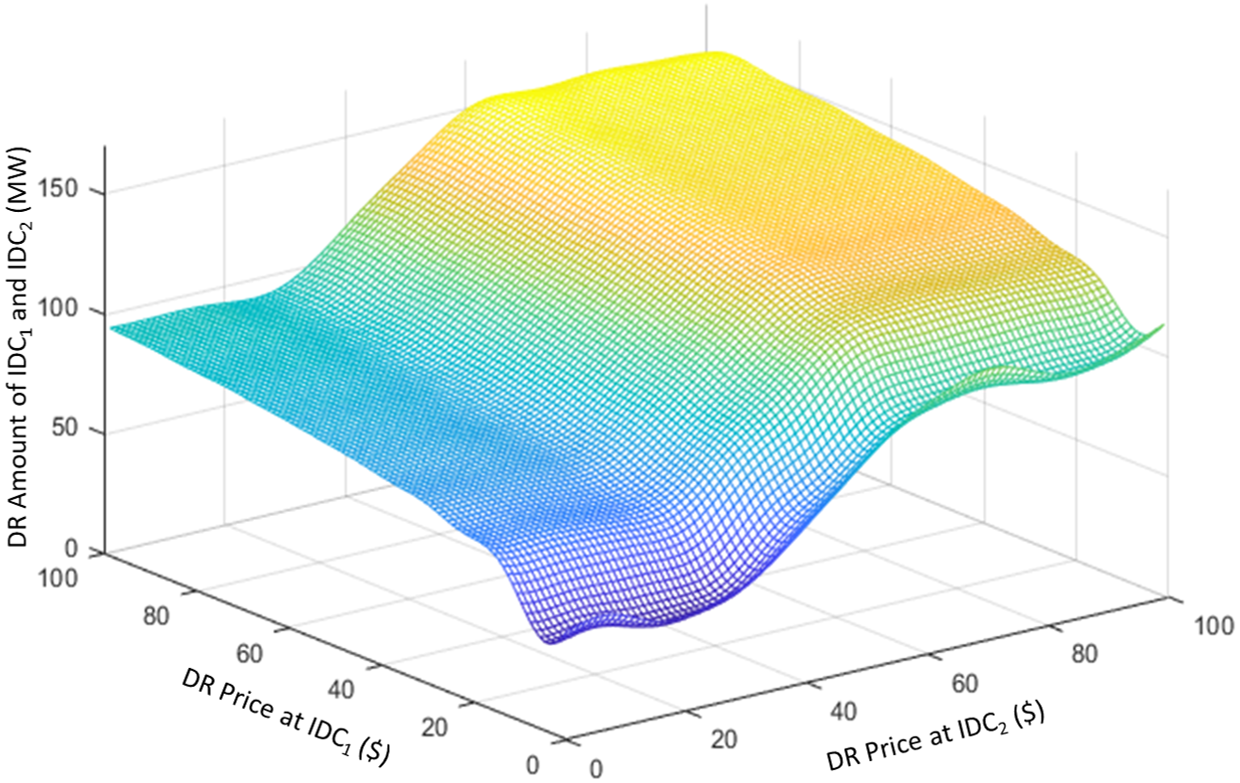} \caption{DR Price-Amount Curve of Multiple IDCs at One Time Slot} 
\label{fig2} 
\end{figure}

Since the dimension number of the estimated function is eight ($P^{DR}=\hat{f}(p^{DR}) \colon\bm{\mathrm{R^8}}\Rightarrow\bm{\mathrm{R^8}}$), the full function cannot be visualized and shown. So the DR Price-Amount curve of two IDCs at one single time slot is illustrated instead as shown in Fig. \ref{fig2}. The x and y axis in Fig. \ref{fig2} is the DR price of two IDCs respectively, and the z axis represent the total DR amount provided by the two IDCs. It can be seen that the dynamic of DR amount increasing with DR price is accurately described by the function. Considering the time-varying workloads in IDCs and the covariation among multiple dimensions, it is impossible to obtain this DR Price-Amount curve based on historical data in conventional ways, and the proposed IDC operation sampling and regressing method solved this issue. Considering the consumed time of generating samples and estimating the function in this case is within 2 minutes, the proposed method satisfies the response time requirement of most DR programs and can be used for demand side management in operational level.

\begin{figure}[ht]
\centering 
\includegraphics[width=0.45\textwidth]{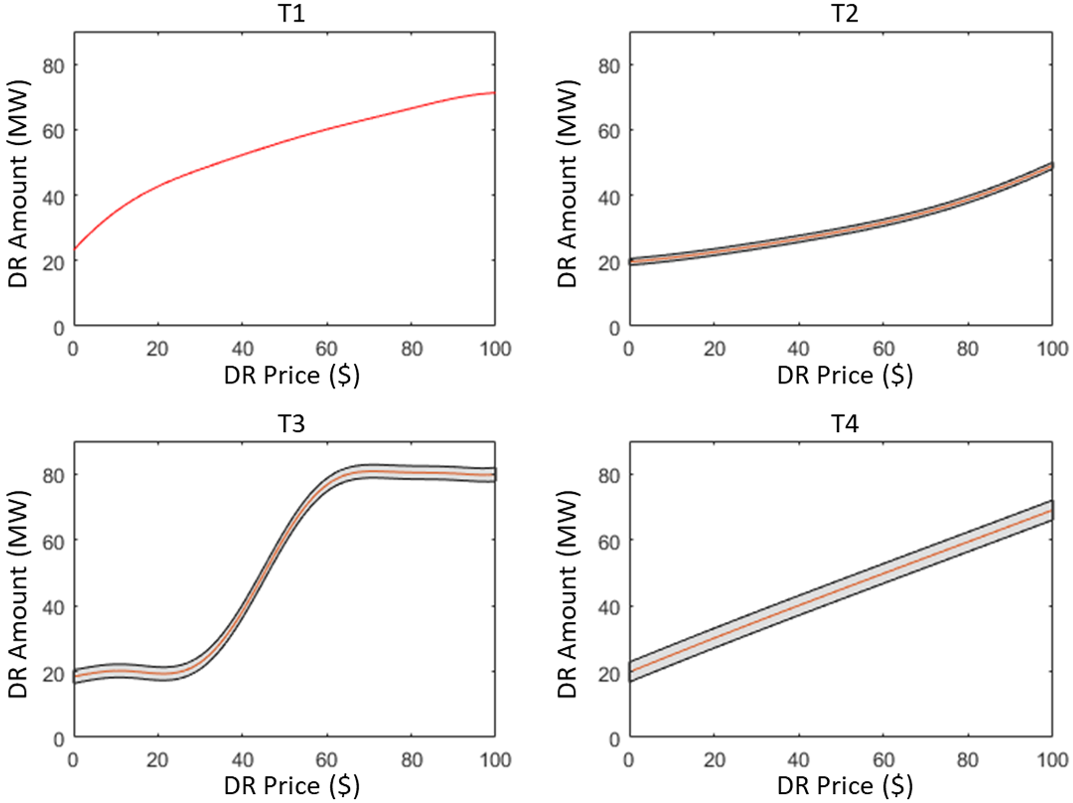} \caption{DR Price-Amount Curves of One IDCs at Multiple Time Slots} 
\label{fig3} 
\end{figure}

Figure \ref{fig3} illustrates the estimated function in another perspective: the DR Price-Amount curves of one IDC in different time slots are shown. The red line in the figures represents the estimated function, and the dark area shows the varying range of the estimation results (95\% confidence level). It can be seen that the time varying uncertaintes in IDC operation shown in Fig. \ref{prediction} is reflected in the estimated function, which will assist in the operation of the power system side. 

As described previously, the selection of kernel functions will affect the estimation accuracy of GPR. So the mainstream kernels are tested here, and the estimation error for data within and out of samples is shown in Tab. \ref{table1}. It can be seen that all kernels except the linear one achieve nearly the same estimation accuracy, which is very similar to the Benchmark 1 (regression tree) and Benchmark 2 (neural network), proving that the estimation error is not because of the estimation method but is caused by the uncertainties in IDC operation.

\begin{table}[]
\centering
\caption{Performance of GPR with Different Kernel Functions}
\label{table1}
\begin{tabular}{cccccc}
\toprule
Method      & \multicolumn{2}{c}{Error} & Method                                                       & \multicolumn{2}{c}{Error} \\
\midrule
\begin{tabular}[c]{@{}c@{}}Kernel\\ Function\end{tabular} &
  \begin{tabular}[c]{@{}c@{}}Within\\ Sample\end{tabular} &
  \begin{tabular}[c]{@{}c@{}}Out of\\ Sample\end{tabular} &
  \begin{tabular}[c]{@{}c@{}}Kernel\\ Function\end{tabular} &
  \begin{tabular}[c]{@{}c@{}}Within\\ Sample\end{tabular} &
  \begin{tabular}[c]{@{}c@{}}Out of\\ Sample\end{tabular} \\
\midrule
SE          & 2.62\%      & 3.13\%      & Matern 3/2                                                   & 2.61\%      & 3.12\%      \\
Exponential & 2.59\%      & 3.11\%      & Matern 5/2                                                   & 2.61\%      & 3.12\%      \\
Linear      & 12.58\%     & 13.26\%     & \begin{tabular}[c]{@{}c@{}}Rational\\ Quadratic\end{tabular} & 2.61\%      & 3.11\%      \\
\midrule
Benchmark 1 & 2.58\%      & 3.10\%      & Benchmark 2                                                  & 0\%         & 3.86\%    
\\ \bottomrule 
\end{tabular}
\end{table}

\section{Conclusion}
This letter proposes a novel method that estimating the DR Price-Amount curve for IDCs. By sampling the IDC operation states, the dynamic of the provided DR amount varing with DR price and the uncertainties in IDC operation are captured. Based on the sampled data, the DR Price-Amount function is estimated by GPR, which can not only accurately describe the DR Price-Amount dynamic, but also provide the varying range of the curve. The proposed method will benefit the power systems by providing more information about the covariation among the DR services provided by multiple IDCs and the uncertainties in the provided DR services, which is crucial for demand side management. The performance of the proposed method are proved by the numerical results. The future research will focus on the integration of the DR Price-Amount curve in networks' optimal operation. 

\bibliographystyle{IEEEtran}
\bibliography{main.bbl}

\end{document}